\documentclass[]{aastex631}

\usepackage{bm}
\usepackage{amsmath}

\newcommand\ltsima{$\; \buildrel <\over\sim \;$}
\newcommand\simlt{\lower.5ex\hbox{\ltsima}}
\newcommand\gtsima{$\; \buildrel >\over\sim \;$}
\newcommand\simgt{\lower.5ex\hbox{\gtsima}}

\shorttitle{Dependence of Planet Frequency}
\shortauthors{Nunota et al.}

\graphicspath{{./}{figures/}}

\defcitealias{suz16}{S16}
\defcitealias{kos23}{K23}

\begin{document}

\title{Measurement of Dependence of Microlensing Planet Frequency on The Host Star Mass and Galactocentric Distance by using a Galactic Model}


\author{Kansuke Nunota}
\affiliation{Department of Earth and Space Science, Graduate School of Science, Osaka University, Toyonaka, Osaka 560-0043, Japan}
\affiliation{Center for Astrophysics $\lvert$ Harvard $\And$ Smithsonian, 60 Garden Street, Cambridge, MA 02138, USA}

\author{Naoki Koshimoto}
\affiliation{Department of Earth and Space Science, Graduate School of Science, Osaka University, Toyonaka, Osaka 560-0043, Japan}

\author{Daisuke Suzuki}
\affiliation{Department of Earth and Space Science, Graduate School of Science, Osaka University, Toyonaka, Osaka 560-0043, Japan}

\author{Takahiro Sumi}
\affiliation{Department of Earth and Space Science, Graduate School of Science, Osaka University, Toyonaka, Osaka 560-0043, Japan}

\author{David P.~Bennett}
\affiliation{Code 667, NASA Goddard Space Flight Center, Greenbelt, MD 20771, USA}
\affiliation{Department of Astronomy, University of Maryland, College Park, MD 20742, USA}
\affiliation{Center for Research and Exploration in Space Science and Technology, NASA/GSFC, Greenbelt, MD 20771, USA}

\author{Aparna Bhattacharya}
\affiliation{Code 667, NASA Goddard Space Flight Center, Greenbelt, MD 20771, USA}
\affiliation{Department of Astronomy, University of Maryland, College Park, MD 20742, USA}
\affiliation{Center for Research and Exploration in Space Science and Technology, NASA/GSFC, Greenbelt, MD 20771, USA}

\author{Yuki Hirao}
\affiliation{Institute of Astronomy, Graduate School of Science, The University of Tokyo, 2-21-1 Osawa, Mitaka, Tokyo 181-0015, Japan}

\author{Sean K. Terry}
\affiliation{Code 667, NASA Goddard Space Flight Center, Greenbelt, MD 20771, USA}
\affiliation{Department of Astronomy, University of Maryland, College Park, MD 20742, USA}

\author{Aikaterini Vandorou}
\affiliation{Code 667, NASA Goddard Space Flight Center, Greenbelt, MD 20771, USA}
\affiliation{Department of Astronomy, University of Maryland, College Park, MD 20742, USA}
\affiliation{Center for Research and Exploration in Space Science and Technology, NASA/GSFC, Greenbelt, MD 20771, USA}

\begin{abstract}
We measure the dependence of planet frequency on host star mass, $M_{\rm L}$, and distance from the Galactic center, $R_{\rm L}$, using a sample of planets discovered by gravitational microlensing.
We compare the two-dimensional distribution of the lens-source proper motion, $\mu_{\rm rel}$, and the Einstein radius crossing time, $t_{\rm E}$, measured for 22 planetary events from \citet{suz16} with the distribution expected from Galactic model.
Assuming that the planet-hosting probability of a star is proportional to $M_{\rm L}^m R_{\rm L}^r$, we calculate the likelihood distribution of $(m,r)$. We estimate that $r = 0.10^{+0.51}_{-0.37}$ and $m = 0.50^{+0.90}_{-0.70}$ under the assumption that the planet-hosting probability is independent of the mass ratio.
We also divide the planet sample into subsamples based on their mass ratio, $q$, and estimate that $m=-0.08^{+0.95}_{-0.65}$ for $q < 10^{-3}$ and $1.25^{+1.07}_{-1.14}$ for $q > 10^{-3}$.
Although uncertainties are still large, this result implies a possibility that in orbits beyond the snowline, massive planets are more likely to exist around more massive stars whereas low-mass planets exist regardless of their host star mass.
\end{abstract}


\section{Introduction} \label{intro}
More than 5500 planets have been discovered to date and gravitational microlensing is one of the most effective methods to detect planets.
Gravitational microlensing is a unique method that can detect planets residing in a wide range of parameter space, such as planets in the Galactic disk \citep{gau08, ben10} or bulge \citep{bha21}, planets around late M-dwarfs \citep[][S. K. Terry et al. in prep.]{ben08} or G-dwarfs \citep{bea16}, and even planets around white dwarfs \citep{bla21}. Measuring the planet frequency as a function of host star mass and location in our Galaxy via microlensing enables us to study the comprehensive picture of planet formation throughout our Galaxy. However, there is a difficulty in determining mass and distance in the microlensing method.\par

For most planetary events, the angular Einstein radius,  $\theta_{\rm E}$, and Einstein radius crossing time, $t_{\rm E}$,  can be measured via light curve analysis as informative parameters of the host star. These parameters are related by the following equations:
\begin{eqnarray} \label{def-tE}
t_{\rm E} = \frac{1}{\mu_{\rm rel}}\sqrt{\kappa M_{\rm L} \left(\frac{\rm 1~au}{D_{\rm L}} - \frac{\rm 1~au}{D_{\rm S}}\right)},
\end{eqnarray}
\begin{eqnarray}\label{def-thetaE}
\theta_{\rm E} = t_{\rm E} \times \mu_{\rm rel},
\end{eqnarray}
where $\kappa = 8.144 ~{\rm mas} ~M_\odot ^{-1}$, $D_{\rm L}$ and $D_{\rm S}$ are distance to the lens and source, respectively, and $M_{\rm L}$ is the lens mass. The lens-source relative proper motion $\mu_{\rm rel}$ is given by $\mu_{\rm rel} = |{\bm \mu_{\rm L}} - {\bm \mu_{\rm S}}|$ where ${\bm \mu_{\rm L}}$ is the lens proper motion vector, and ${\bm \mu_{\rm S}}$ is the source proper motion vector. It is clear from these equations that the two parameters $t_{\rm E}$ and $\theta_{\rm E}$ alone cannot determine $M_{\rm L}$ and $D_{\rm L}$, even assuming that the source star is located in the Galactic bulge (i.e., $D_{\rm S} \sim 8~$kpc). Therefore, to determine the lens mass and distance, it is necessary to measure at least one of the additional quantities that determine the mass--distance relations: microlens parallax or lens brightness.
However, there are too few planetary events with measured microlens parallax to obtain statistically useful constraints, since the microlens parallax signal is usually subtle and it is mostly difficult to detect such a signal with ground-based surveys.
Also, the lens brightness measurements require high-angular-resolution follow-up observations several years after the event \citep{bha21, bla21}, making it difficult to obtain sufficient statistics at this moment.\par

Due to these difficulties, the dependence of planetary frequency on the host star mass and the location in our Galaxy is not yet well understood.
\citet{kos21a} attempted to measure the dependence of planet frequency on both host star mass ($\sim M_{\rm L}$) and on the Galactocentric distance ($R_{\rm L}$) by assuming the planet-hosting probability $P_{\rm host} \propto M_{\rm L}^m R_{\rm L}^r$. 
They have compared the $\mu_{\rm rel}$ distribution for given $t_{\rm E}$ of 28 planetary events by \citet{suz16}, \citet{gou10} and \citet{cas12} with the distribution expected by a Galactic model to estimate $m$ and $r$. 
They estimated $r = 0.2 \pm 0.4$ and concluded that there is no large dependence of planet frequency on Galactocentric distance. 
However, their estimate for the parameter of the dependence on host mass was highly uncertain, $m= 0.2 \pm 1.0$.
The large uncertainty in $m$ is partly because they used the distribution of $\mu_{\rm rel}$ given $t_{\rm E}$ instead of the distribution of $t_{\rm E}$ and $\mu_{\rm rel}$.
This contrivance enabled them to avoid detection efficiency calculations but corresponded to a reduction of the two-dimensional information contained in the original distribution of $t_{\rm E}$ and $\mu_{\rm rel}$ to the one-dimensional information contained in the distribution of $\mu_{\rm rel}$ given $t_{\rm E}$.
This in turn means that the two-dimensional distribution of $t_{\rm E}$ and $\mu_{\rm rel}$ can be used to further constrain $m$ and $r$, as long as the detection efficiency is available.

Recently, \citet{kos23} (hereafter, 
\citetalias{kos23}) calculated the detection efficiency for single lens events for the MOA-II 9-yr survey by image-level simulations.
This study utilizes their image-level simulations combined with the detection efficiency for {\it planetary signals} by \citet{suz16} (hereafter, \citetalias{suz16}) to calculate the detection efficiency for {\it planetary events} of the \citetalias{suz16} sample.
By using this combined detection efficiency, we
compare the $(t_{\rm E},\mu_{\rm rel})$ distribution of the MOA-II  planet sample \citepalias{suz16} with the predicted one from
the Galactic model optimized toward the Galactic bulge
\citep{kos21b} to estimate $m$ and $r$.


This paper is organized as follows.
We describe our method in Section \ref{sec-method}.
Section \ref{sec-planet} presents the analysis for the \citetalias{suz16}'s planetary event sample to calculate the likelihood distribution of $(m, r)$. Discussions are presented in Section \ref{sec-dis} and Section \ref{sec-conclu} contains the conclusions.

\section{Method} \label{sec-method}
We follow the method of \citet{kos21a} except that we do not give $t_{\rm E}$ as a fixed value and consider detection efficiency instead.
The main objective of this study is to estimate a dependence of planet frequency on the host star mass and the Galactocentric distance by comparing the ($t_{\rm E},\mu_{\rm rel}$) distribution observed in planetary microlensing events with that distribution predicted from a Galactic model. In this paper, a Galactic model refers to a combination of stellar mass function, stellar density and
velocity distributions in our Galaxy, which enables us to calculate the microlensing event rate $\Gamma$ as a function of microlensing parameters.

We denote the parameter distribution of microlensing events expected from a Galactic model as $\Gamma_{\rm all} \left(t_{\rm E}, \mu_{\rm rel}, M_{\rm L}, R_{\rm L}\right)$. 
Note that $\Gamma_{\rm all}$ represents the parameter distribution for all microlensing events, regardless of whether each system has a planet or not, or whether each microlensing events are detected.
If we assume that the planet-hosting probability is proportional to $M_{\rm L}^m R_{\rm L}^r$, 
the ($t_{\rm E},\mu_{\rm rel}$) distribution for planetary events, $\Gamma_{\rm host}$, is given by
\begin{equation}
\Gamma_{\rm host}\left(t_{\rm E},\mu_{\rm rel}|m,r\right) \propto
\int dM_{\rm L}dR_{\rm L} \, \Gamma_{\rm all}\left(t_{\rm E}, \mu_{\rm rel}, M_{\rm L}, R_{\rm L}\right) M_{\rm L}^m R_{\rm L}^r\, .
\label{gammma_p} 
\end{equation}
Then, the probability of observing a planetary event with $({t_{\rm E}^{\rm(obs)}, \mu_{\rm rel}^{\rm (obs)}})$ is given by
\begin{equation}
f\left({t_{\rm E}^{\rm (obs)}, \mu_{\rm rel}^{\rm (obs)}}|m,r\right) =
\int dt_{\rm E}^{,}d\mu_{\rm rel}^{,} \Bigl[ k\left({t_{\rm E}^{\rm (obs)}, \mu_{\rm rel}^{\rm (obs)}};t_{\rm E}^{,},\mu_{\rm rel}^{,}\right)
\Gamma_{\rm host}\left(t_{\rm E}^{,}, \mu_{\rm rel}^{,}|m,r\right) \epsilon\left(t_{\rm E}^{,}, \mu_{\rm rel}^{,}\right)\Bigr] ,
\label{eq-fobs} 
\end{equation}
where $\epsilon(t_{\rm E}, \mu_{\rm rel})$ is the detection efficiency for a planetary event. Note that the dependence of detection efficiency on $\mu_{\rm rel}$ is negligible for the \citetalias{suz16} planetary event sample as discussed in Section \ref{planet-det-eff}.
$k(t_{\rm E}^{\rm (obs)}, \mu_{\rm rel}^{\rm (obs)};t_{\rm E}^{,},\mu_{\rm rel}^{,})$ is a kernel function, and we adopt a Gaussian kernel,
\begin{equation}
k({t_{\rm E}^{\rm (obs)}, \mu_{\rm rel}^{\rm (obs)}};t_{\rm E}^{,},\mu_{\rm rel}^{,}) =
\frac{1}{\sqrt{2\pi \sigma_{t_{\rm E}}^{\left(obs\right)2}}}\exp\left(-\frac{(t_{\rm E}^{\rm (obs)} - t_{\rm E}^{,})^2}{2 \sigma_{t_{\rm E}}^{\rm (obs)2}} \right)
\frac{1}{\sqrt{2\pi \sigma_{\mu_{\rm rel}}^{\left(obs\right)2}}} \exp \left(-\frac{(\mu_{\rm rel}^{\rm (obs)} - \mu_{\rm rel}^{,})^2}{2 \sigma_{\mu_{\rm rel}}^{\left(obs\right)2}}\right),
\label{kernel}
\end{equation}
where $\sigma_{t_{\rm E}}^{\rm (obs)}$ and $\sigma_{\mu_{\rm rel}}^{\rm (obs)}$ are the uncertainty of ${t_{\rm E}}^{\rm (obs)}$ and $\mu_{\rm rel}^{\rm (obs)}$ respectively. The introduction of a kernel function is intended to allow some uncertainty in the observed values.

Note that we use the parameter set of $(t_{\rm E},\mu_{\rm rel})$ rather than $(t_{\rm E},\theta_{\rm E})$ because $\mu_{\rm rel}$ is less correlated with $t_{\rm E}$ than $\theta_{\rm E}$ and usually has a smaller error bar when it is determined via the finite source effect \citep{alc97, yoo04}.
The angular Einstein radius and the relative proper motion are represented by $\theta_{\rm E} = \theta_*/\rho$ and $\mu_{\rm rel} = \theta_*/(t_{\rm E} \, \rho)$, respectively, with the finite source parameter $\rho$ and the angular source radius $\theta_*$.
While $t_{\rm E} \, \rho$\,---which is often defined as the source radius crossing time $t_*$---\, is well determined by the light curve as also suggested by \citet{yee12}, $\rho$ tends to be anti-correlated with $t_{\rm E}$.
Thus, the correlation between $t_{\rm E}$ and $\mu_{\rm rel}$ is smaller than the one between $t_{\rm E}$ and $\theta_{\rm E}$.
Nevertheless, we performed the same analysis by using the parameter set of $(t_{\rm E},\theta_{\rm E})$ and confirmed that our results would not change significantly.

When a sample of $N_{\rm sample}$ events is given, the probability of observing those events under a specific combination of $(m,r)$, ${\cal L}(m,r)$, is expressed as
\begin{eqnarray}
{\cal L} (m, r) = \prod_{i = 1}^{N_{\rm sample}} f(t_{{\rm E},i}^{\rm (obs)}, \mu_{{\rm rel},i}^{\rm (obs)}| m,r).
\label{eq-L}
\end{eqnarray}
By calculating Eq. (\ref{eq-L}) under various values of $(m,r)$ and comparing the values of ${\cal L} (m, r)$, it is possible to evaluate which $(m,r)$ values are more likely.
In this paper, we calculate ${\cal L} (m, r)$ in a grid of $0.2$ increments in the range of $-3 \leq m \leq 3$ and $-3 \leq r \leq 3$.
This corresponds to applying a uniform prior distribution of -3 to  3 for $m$ and $r$ and calculating the posterior probability distribution.

In this study, we use the Galactic model developed by \citet{kos21b} and their microlensing event simulation tool, \texttt{genulens}\footnote{\url{https://github.com/nkoshimoto/genulens}} \citep{kos22}.
This model was designed to reproduce the stellar distribution toward the Galactic bulge by fitting to the Gaia DR2 velocity data \citep{kat18}, OGLE-III red clump star count data \citep{nat13}, VIRAC proper motion catalog \citep{smi18, cla19}, BRAVA radial velocity measurements \citep{ric07, kun12}, and OGLE-IV star count and microlensing rate data \citep{mro17, mro19}.
The stellar mass considered in this model ranges from $10^{-3} ~ M_\odot$ to $5.3 ~ M_\odot$ and the typical lens mass ranges from $0.02 ~ M_\odot$ to $1.0 ~ M_\odot$.
Although the model is optimized for a microlensing study toward the Galactic bulge, we would like to ensure that no significant bias is introduced in our result by the model since our analysis strongly depends on the Galactic model used.


To validate the approach of this study, we present two types of analysis in Appendices. 
First, Appendix \ref{sec-mock-test} presents mock data analysis, where we generate 50 artificial planetary events based on the Galactic model with certain $(m, r)$ values and calculate the likelihood distributions. 
As a result, we confirmed this method can reproduce the input $(m, r)$ values properly.
Then, Appendix \ref{sec-model} compares the $(t_{\rm E}, \mu_{\rm rel})$ distribution of the MOA-II 9-yr FSPL sample with the distribution predicted by the Galactic model. This confirmed no significant bias would be introduced in our result by using the Galactic model by \citet{kos21b}.

\begin{figure*}
\centering
\includegraphics[width=15cm]{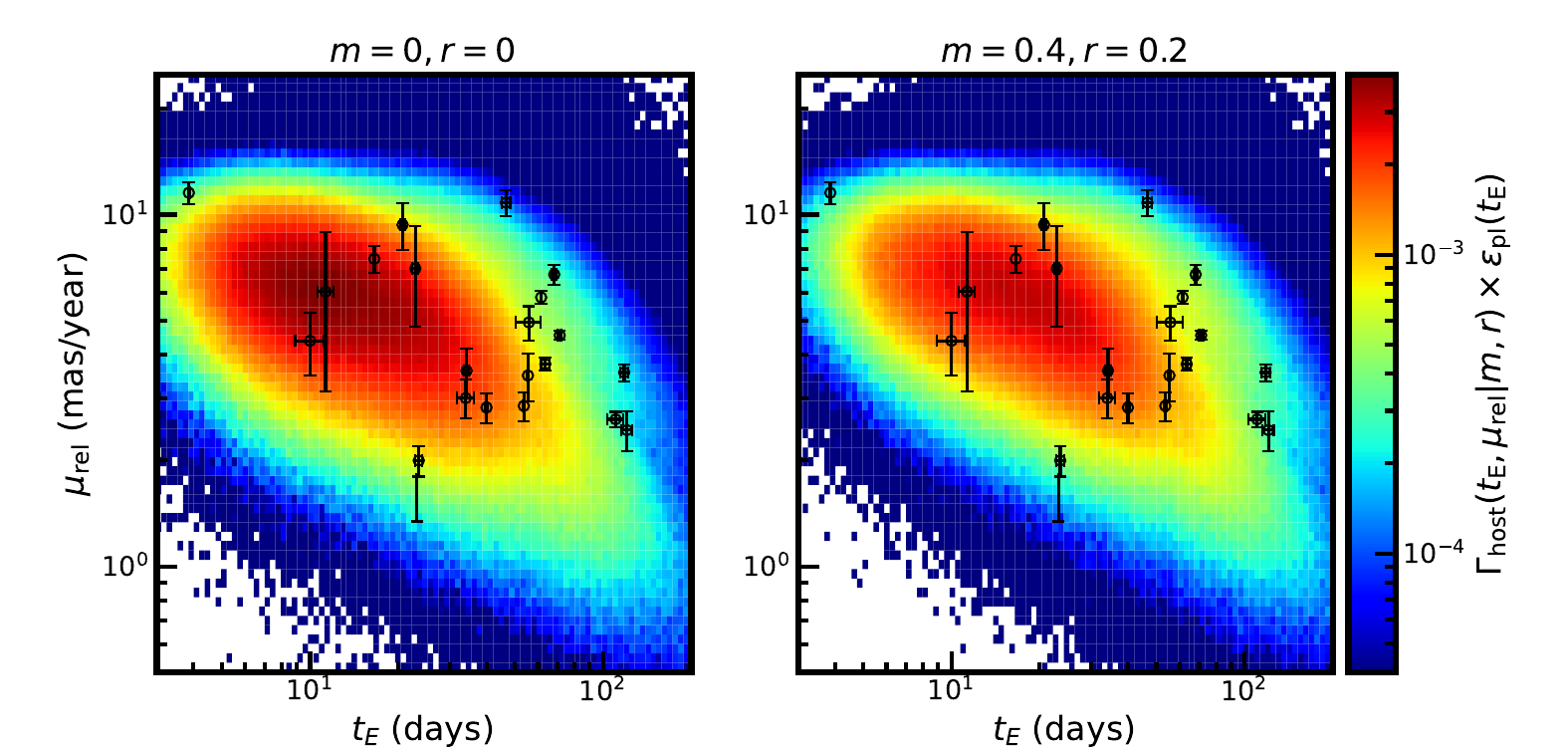}
\caption{
Comparison of the $(t_{\rm E}, \mu_{\rm rel})$ distributions for the planetary event sample of \citetalias{suz16} between the observations (black points) and the prediction by the Galactic model combined with the detection efficiency (color maps).
The color map in the left panel shows the distribution predicted by the model with $P_{\rm host} \propto {\rm const}$, i.e., assuming that all stars are equally likely to host planets, regardless of their mass or location in our Galaxy.
The color map in the right panel is the distribution predicted by the model with $P_{\rm host} \propto M_{\rm L}^{0.4}R_{\rm L}^{0.2}$, corresponding to the grid that gives the maximum likelihood. 
}
\label{fig-tE_murel_planet}
\end{figure*}

\section{Application}\label{sec-planet}

\subsection{Planetary Microlensing Event Sample} \label{sec-pla_sample}

\citet{kos21a} used 28 planetary events consisting of 22 events from the MOA-II survey during 2007--2012\footnote{The \citetalias{suz16} original sample consists of 23 events. However, the ambiguous event OGLE-2011-BLG-0950  turned out to be a stellar binary event \citep{ter22}.} \citepalias{suz16} and 6 additional events from \citet{gou10} and \citet{cas12}.
The mixture of samples from different surveys was valid because they did not need to consider the detection efficiency by focusing on the one-dimensional distribution of $\mu_{\rm rel}$ for given $t_{\rm E}$.
However, our analysis, which focuses on the two-dimensional distribution of $t_{\rm E}$ and $\mu_{\rm rel}$, requires consideration of detection efficiency as described in Section \ref{sec-method}.
Because the detection efficiency used in this analysis is optimized for the MOA-II survey as described in Section \ref{planet-det-eff}, we only use the 22 planetary events from the MOA-II survey as our sample in this study.

\citet{zhu14} predicted that 55\% of planets would be detected without caustic crossing for a high-cadence microlensing survey the KMTNet sample confirmed this prediction.
On the other hand, 5 events out of our 22 planetary events were detected without caustic crossing.
It indicates a higher percentage of caustic crossings than the prediction by \citet{zhu14}.
This might be due to differences in observation cadence or signal-to-noise ratio between the MOA-II survey and the KMTNet survey.

The black points in Fig. \ref{fig-tE_murel_planet} show the ($t_{\rm E}$, $\mu_{\rm rel}$) distribution of the 22 planetary events.
These values are taken from the discovery or follow-up papers of each event.
Some notes are as follows:
MOA-2011-BLG-322 \citep{shv14} has only a lower limit on $\mu_{\rm rel}$; 
MOA-2011-BLG-262 \citep{ben14} has the fast and slow solutions, but we use only the slow solution because it has a much larger prior probability as discussed in \citet{ben14}. The $t_{\rm E}$ and $\mu_{\rm rel}$ values of the following events are from papers in preparation regarding follow-up high-angular resolution imaging:
MOA-2007-BLG-192 (S. K. Terry et al. in prep.), MOA-2010-BLG-328 (A. Vandorou et al. in prep.), and OGLE-2012-BLG-0563 (A. Bhattacharya et al. in prep.).

Although the following analysis will be performed with the updated values taken from the papers in preparation, we have also performed the same analysis with the values from the original discovery papers and confirmed that our results would not change significantly.

\subsection{Detection Efficiency} \label{planet-det-eff}
As discussed in Section \ref{sec-method}, our method requires the detection efficiency for the \citetalias{suz16}'s planetary event sample.
When calculating the detection efficiency of a sample, we need to carefully consider how the sample was collected.
The event selection process of \citetalias{suz16} can be interpreted by the following two steps; 
(i) the 1474 ``well-monitored" events were selected from the 3300 events alerted by the MOA group during 2007--2012 based on the selection criteria summarized in Table 1 of \citetalias{suz16}, and 
(ii) the 22 planetary and 1 ambiguous events were selected among them based on the $\chi^2$ difference between the single-lens model and the planetary model.
Therefore, we represent the detection efficiency for the \citetalias{suz16}'s planetary event sample by
\begin{align}
\epsilon_{\rm pl} (t_{\rm E}) = \epsilon_{\rm WM} (t_{\rm E}) \, \epsilon_{\rm ano} (t_{\rm E}), \label{eq:ep_pl}
\end{align}
where $\epsilon_{\rm WM}$ is the detection efficiency for the well-monitored events in \citetalias{suz16} and $\epsilon_{\rm ano}$ is the detection efficiency for the planetary anomaly feature.

The detection efficiency for the planetary anomaly feature was calculated by \citetalias{suz16}, and 
we use the data shown in Figure 10 of \citetalias{suz16} as $\epsilon_{\rm ano} (t_{\rm E})$.
The dependence of the detection efficiency on $t_{\rm E}$ changes according to the mass ratio.
Therefore, we use different detection efficiencies corresponding to the mass ratio for each event.
In principle, the detection efficiency for a planetary anomaly feature depends not only on $t_{\rm E}$ but also on $\mu_{\rm rel}$ through the source radius crossing time, $t_* = \theta_*/\mu_{\rm rel}$, where $\theta_*$ is the angular source radius.
However, as discussed in \citet{kos21a}, this effect can be considered negligible in the current case because the dependence of the detection efficiency for the anomaly feature on $t_*$ is 
negligibly small for a mass ratio of $q > 10^{-4}$ \citepalias[see Figure 7 of][]{suz16}, which dominates our sample.

\begin{figure}
\centering
\includegraphics[width=9cm]{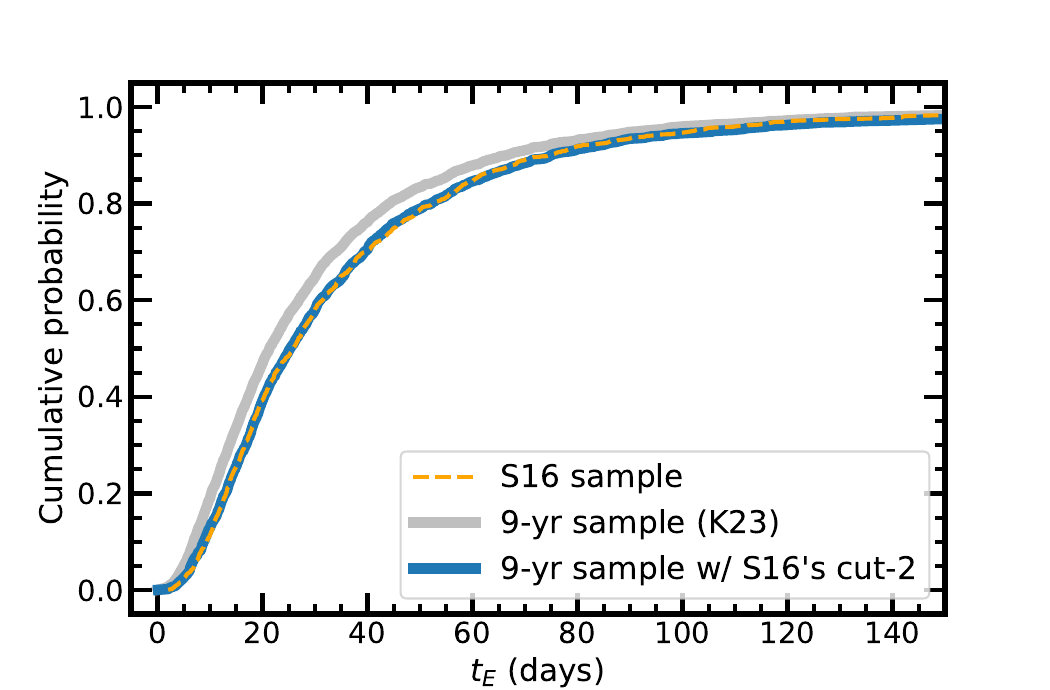}
\caption{
Comparison of the cumulative distributions of the Einstein radius crossing time, $t_{\rm E}$, among the three samples; the \citetalias{suz16} sample (orange dashed line), the MOA-II 9-yr sample (gray solid line), and the 9-yr sample re-selected by adding the \citetalias{suz16}'s cut-2 criteria (blue solid line).
}
\label{cumu_tE}
\end{figure}

On the other hand, \citetalias{suz16} did not calculate the detection efficiency for the well-monitored events, $\epsilon_{\rm WM}$, because they did not use the $t_{\rm E}$ information for their analysis and $\epsilon_{\rm WM}(t_{\rm E})$ was not needed.
The well-monitored events were selected from the events alerted by the MOA alert system, which depends on the observer who was monitoring the light curve of each microlens candidate at the time.
Although it is difficult to reproduce the exact selection process, we here utilize the results of the image-level simulation of the $6.4 \times 10^7$ artificial events conducted by \citetalias{kos23} for their MOA-II 9-yr analysis to estimate $\epsilon_{\rm WM}$.
In Fig. \ref{cumu_tE}, the orange dashed line and gray line show the cumulative $t_{\rm E}$ distributions for the \citetalias{suz16} sample and 9-yr sample, respectively. 
We can see a lack of short timescale events in the \citetalias{suz16}'s sample compared to the 9-yr sample.
This is expected because \citetalias{kos23} studied free-floating planets, which have very short timescales, by selecting all events including short timescale events, whereas \citetalias{suz16} selected only well-monitored events which preferentially have longer timescales.
To make the $t_{\rm E}$ distribution of the 9-yr sample closer to the \citetalias{suz16}'s one, we added the \citetalias{suz16}'s cut-2 criteria, i.e., $\sigma_{u_0}/u_0 < 0.40$ or $\sigma_{u_0} < 0.02$ and $\sigma_{t_{\rm E}}/t_{\rm E} < 0.25$ and $\sigma_{t_{\rm E}} < 20~{\rm days}$, to the original selection process of the 9-yr sample.
The blue solid line in Fig. \ref{cumu_tE} shows the cumulative $t_{\rm E}$ distribution of the re-selected 9-yr sample with the additional \citetalias{suz16}'s cut-2 criteria, which almost perfectly follows the orange dashed line of the \citetalias{suz16}'s sample.
To quantify the similarity, we performed a Kolmogorov-Smirnov test on the two samples and got a $p$-value of $p = 0.949$, which supports the idea that the two distributions were sampled from populations with approximately the same distributions.
This provides a basis for considering that the detection efficiencies of the \citetalias{suz16} sample and the re-selected 9-yr sample are almost the same. 
Therefore, we follow the \citetalias{kos23}'s detection efficiency calculation using their artificial event sample from the image-level simulation with the \citetalias{suz16}'s cut-2 criteria in addition to the original criteria listed in Table 2 of \citetalias{kos23}, and we use it as the detection efficiency for the well-monitored events, $\epsilon_{\rm WM} (t_{\rm E})$.
See Appendix \ref{FSPL_det_eff} for an example calculation of the detection efficiency using the simulated artificial events (and see \citetalias{kos23} for more detail).

\subsection{Likelihood Analysis} \label{sec-result-planet}

\begin{figure}
\centering
\includegraphics[width=9cm]{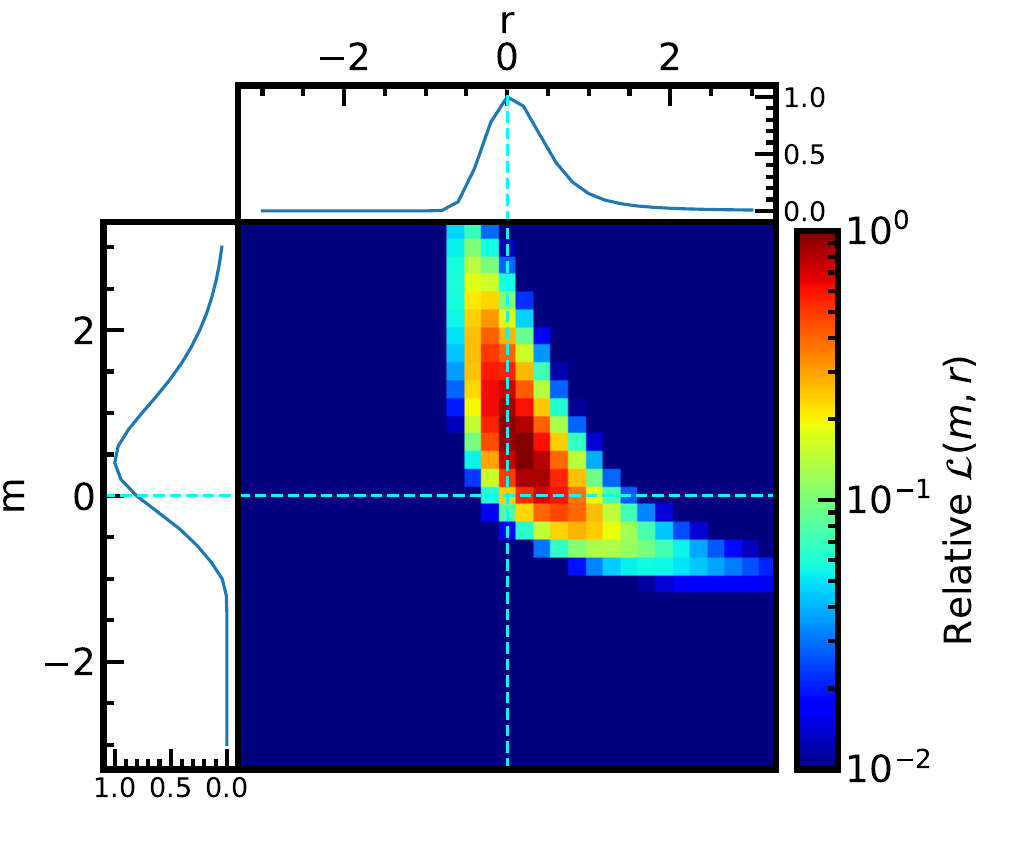}
\caption{
Relative likelihood distribution of $(m, r)$ calculated by Eq. (\ref{eq-L}) for the \citetalias{suz16}'s 22 planetary event sample.
The top panel shows a relative probability distribution of $r$ integrated over $-3< m <3$ uniformly, and the side panel shows a relative probability distribution of $m$ integrated over $-3< r <3$ uniformly.
}
\label{main_result}
\end{figure}

\begin{deluxetable*}{ccccccccc}
\tablewidth{10pt} 
\tablenum{1}
\tablecaption{Result of the likelihood analysis \label{table-result}}
\tablehead{
\colhead{} & \colhead{ all } & \colhead{} & \multicolumn{2}{c}{two-bin} & \colhead{} & \colhead{} & \colhead{three-bin} & \colhead{}\\
\cline{2-2}
\cline{4-5}
\cline{7-9}
\colhead{} & \colhead{$10^{-4.25} < q < 10^{-1.55}$} & \colhead{} & \colhead{$q < 10^{-3}$} &\colhead{$10^{-3}<q$} & \colhead{} & \colhead{$q < 10^{-3.5}$}  & \colhead{$10^{-3.5}<q<10^{-2.5}$} & \colhead{$10^{-2.5}<q$}\\
\colhead{$N_{\rm sample}$}& \colhead{22} & \colhead{} & \colhead{13} & \colhead{9} & \colhead{} & \colhead{6} & \colhead{10} & \colhead{6}
}
\startdata 
$m$ & $0.50^{+0.90}_{-0.70}$ & & $-0.08^{+0.95}_{-0.65}$ & $1.25^{+1.07}_{-1.14}$ &  & $0.46^{+1.29}_{-0.98}$ & $-0.34^{+1.02}_{-0.57}$ & $1.63^{+0.92}_{-1.18}$ \\
$r$ & $0.10^{+0.51}_{-0.37}$ & & $0.41^{+0.95}_{-0.54}$ & $-0.22^{+0.68}_{-0.45}$ &  & $0.29^{+0.98}_{-0.58}$ & $0.76^{+1.22}_{-0.81}$ & $-0.68^{+0.74}_{-0.78}$  
\enddata
\tablecomments{This table shows median and $1~\sigma$ error for each sample.}
\end{deluxetable*}

We calculate the likelihood given by Eq. (\ref{eq-L}) for the 22 planetary events, and 
Fig. \ref{main_result} shows the resulted relative likelihood distribution as a function of $m$ and $r$.
The relative likelihood value at $(m, r) = (0, 0)$ is 0.24 and the relative likelihood takes its maximum value of 1 at $(m,r) = (0.4,0.2)$. 
We find $m = 0.50^{+0.90}_{-0.70}$ and $r = 0.10^{+0.51}_{-0.37}$ from the marginalized distributions.
Our result is consistent with $(m, r) = (0, 0)$, i.e., the idea that all stars are equally likely to host planets.
However, it prefers $m > 0$, suggesting a possible correlation between the planet frequency and the host star mass.
On the other hand, no preference is seen in either positive or negative $r$, which confirms the result of \citet{kos21a} who found no large dependence of planet frequency on the Galactocentric distance.

The color maps of Fig. \ref{fig-tE_murel_planet} show the $(t_{\rm E}, \mu_{\rm rel})$ distributions expected by the model, i.e., $\Gamma_{\rm host}(t_{\rm E},\mu_{\rm rel}|m,r) \times \epsilon_{\rm pl} (t_{\rm E})$, at $(m,r) = (0, 0)$ on the left and $(m,r) = (0.4,0.2)$ on the right. 
Fig. \ref{fig-tE_murel_planet} certainly shows that the expected distribution at the best-fit grid of $(m,r) = (0.4,0.2)$ is more matched with the observational distribution from \citetalias{suz16} than the expected distribution at $(m,r) = (0,0)$.

We also divide the 22-event sample into subsamples by mass ratio and perform the same analysis for these subsamples to see if there is any relationship between the mass ratio of a planetary system and planet frequency.
We try two types of bin patterns for dividing the sample; the first one is the two-bin subsamples with $\log q < -3.0$ (13 events) and $-3.0 < \log q$ (9 events), and the second one is the three-bin subsamples with $\log q < -3.5$ (6 events), $-3.5 < \log q < -2.5$ (10 events), and $-2.5 < \log q$ (6 events).

\begin{figure*}
\centering
\includegraphics[width=10cm]{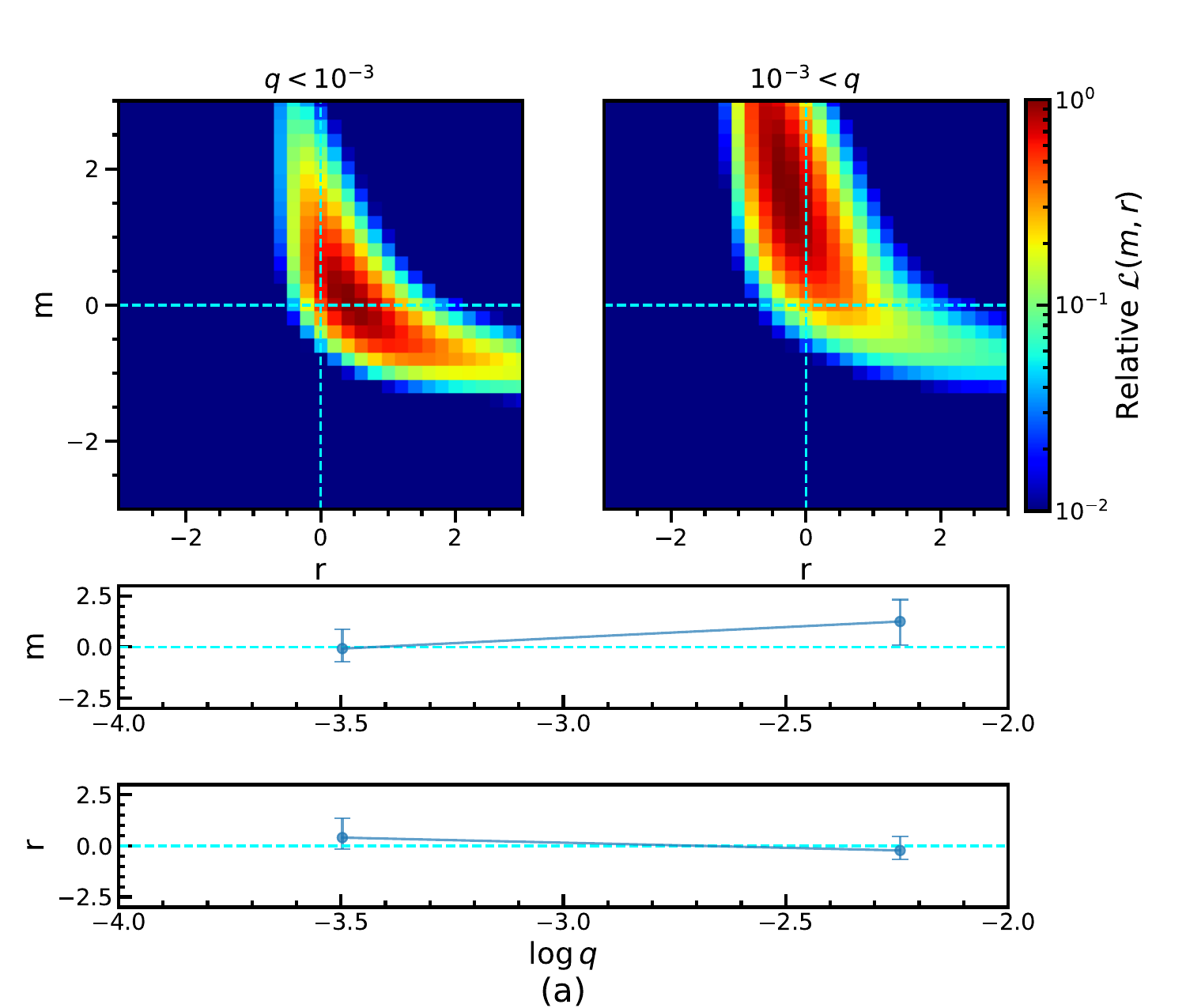}
\includegraphics[width=15cm]{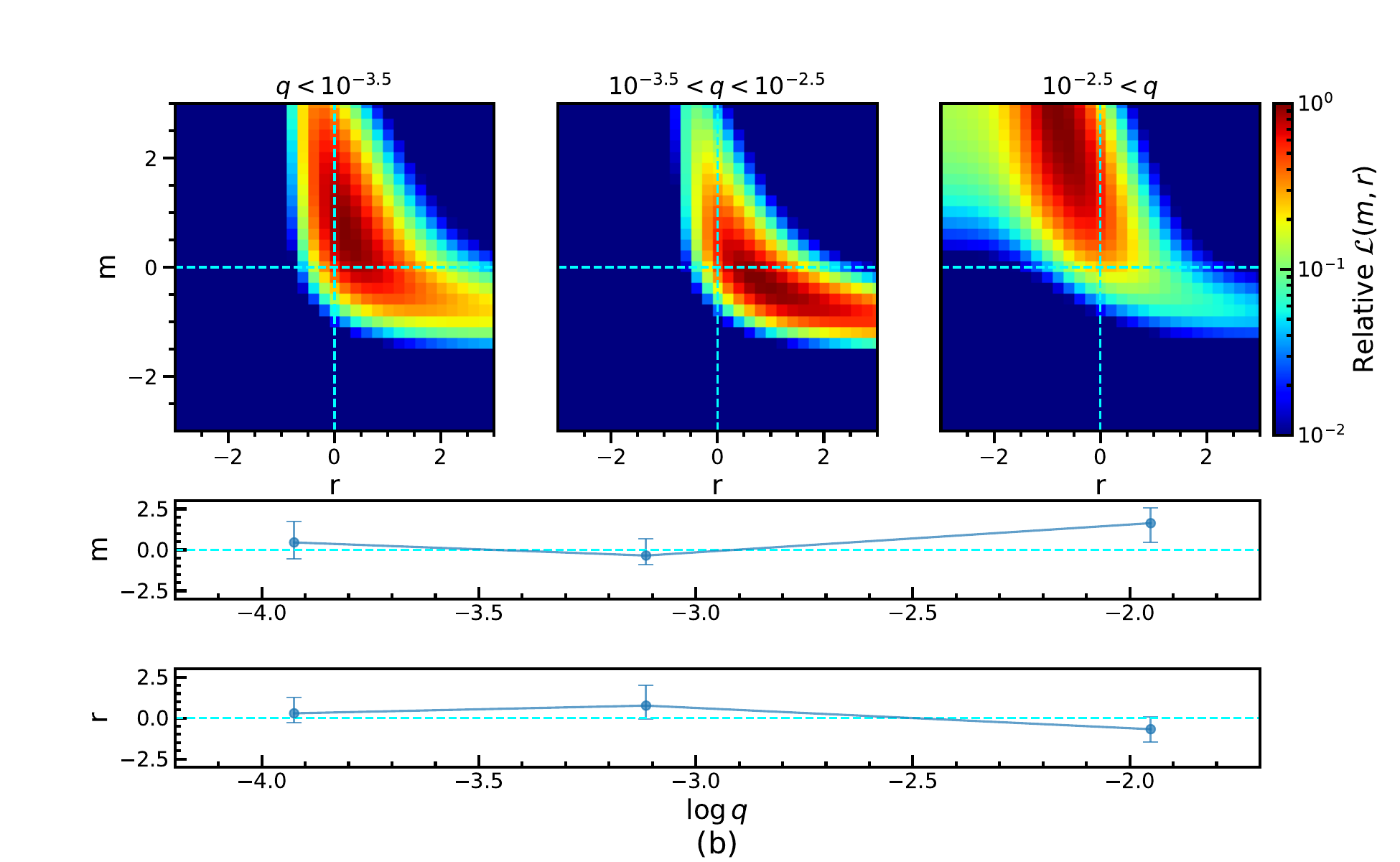}
\caption{
Relative likelihood distributions for (a) the two-bin subsamples and (b) the three-bin subsamples.
In each of (a) and (b) figures, the top panels show the likelihood distribution of $(m, r)$ for the subsample in each bin.
The middle panel shows the median and $1 \sigma$ error of the marginalized $m$ distribution versus the mean of $\log q$ for each bin. The bottom panel is the same for the marginalized $r$ distribution.
}
\label{bin_result}
\end{figure*}

Fig. \ref{bin_result} shows the results of the likelihood analysis for the subsamples, where Fig. \ref{bin_result} (a) is for the two-bin subsamples and Fig. \ref{bin_result} (b) is for the three-bin subsamples.
In each of Figs. \ref{bin_result} (a) and (b), the middle panel plots the mean of $\log q$ vs the median and $1 \sigma$ range of the marginalized $m$ distribution of each subsample, while the bottom panel shows those for the marginalized $r$ values.
Both results show that $m$ is likely to be higher than 0 at the highest $\log q$ bin while $m$ is fully consistent with 0 at the other bins.
This result might suggest that massive planets are more likely to exist around more massive stars whereas low-mass planets are more universal regardless of their host star mass.
On the other hand, the $r$ value seems to have a smaller mass ratio dependence than the $m$ value, although there is an anti-correlation between $m$ and $r$.

While these are potentially interesting features, it is statistically too weak to conclude whether these features are real or not. 
We further discuss the possible dependence of $m$ on the mass ratio in Section \ref{main-discussion}.
The median and $1~\sigma$ values of all likelihood analyses are listed in Table \ref{table-result}.

\section{Discussion} \label{sec-dis}

\begin{figure}
\centering
\includegraphics[width=10cm]{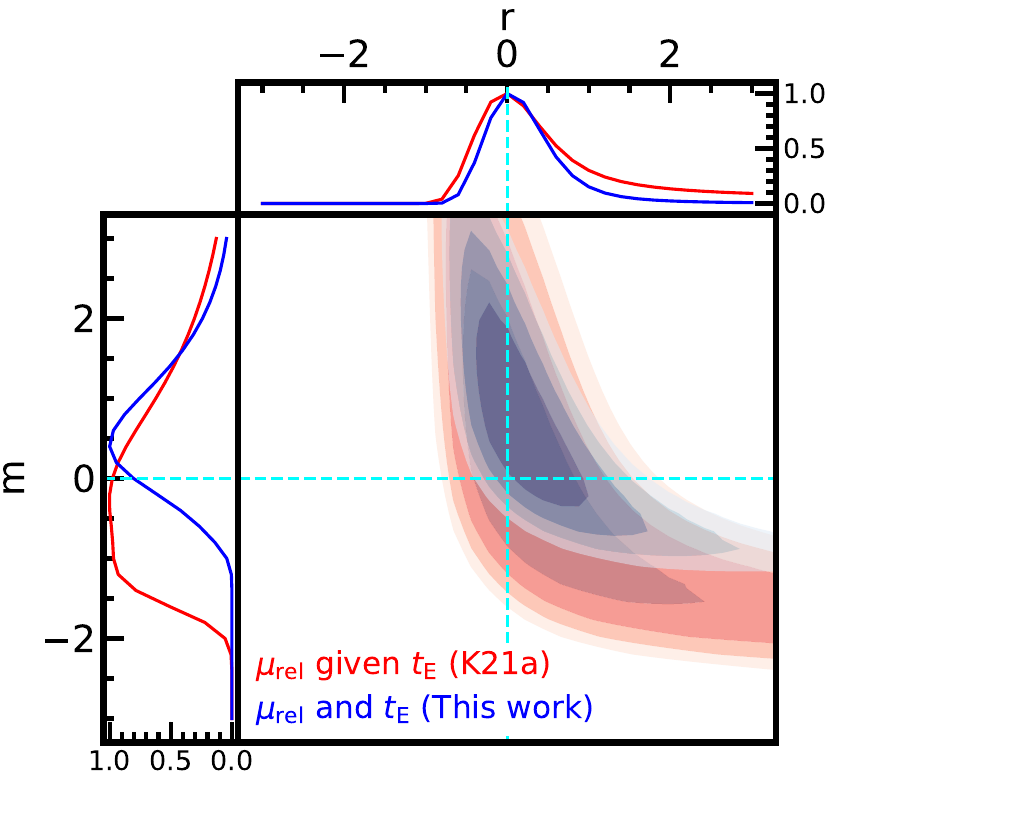}
\caption{
Comparison of the results with \citet{kos21a}'s method (red) that uses one-dimensional $\mu_{\rm rel}$ distributions for given $t_{\rm E}$ and this study's method (blue) that uses two-dimensional $(t_{\rm E}, \mu_{\rm rel})$ distributions.
}
\label{comp_kos_nuno}
\end{figure}

\subsection{Dependence of planet frequency on the host star mass} \label{main-discussion}
We estimated the planet--hosting probability as $P_{\rm host} \propto M_{\rm L}^{0.50^{+0.90}_{-0.70}} \times R_{\rm L}^{0.10^{+0.51}_{-0.37}}$ by using the 22 planetary event sample (Fig. \ref{main_result} and Table \ref{table-result}) for planets beyond the snowline.
Although all the host star masses in our sample have not been measured, the typical mass of the host star is $\sim 0.6~M_{\odot}$ \citepalias{suz16}.
The result that the likelihood distribution prefers $m > 0$ suggests that the planet frequency has a possible positive correlation with the host star mass.
A possible positive correlation is also seen in massive planet subsamples with $q \simgt 10^{-3}$, whereas $m$ is consistent with 0 in lower-mass ratio subsamples.
This implies that giant planets are more likely to exist around more massive stars, whereas lower-mass planets more uniformly exist regardless of the host star mass.

A positive correlation between planet frequency and host star mass, $M_{\rm host}$, for giant planets is also suggested by RV studies for inner planets \citep{jon07,jon10,ref15}.
\citet{jon10} analyzed 1266 stars and estimated that planet frequency is $\propto M_{\rm host}^{1.0\pm 0.3}$.
Their sample ranges from low-mass M-dwarfs with $0.2\,M_{\odot}$ to intermediate-mass subgiants with $1.9\,M_{\odot}$.
\citet{ref15} found that the giant planet frequency increases in the host star mass from $1\, M_{\odot}$ to $1.9\, M_{\odot}$ by analyzing samples from $Lick$ Observatory.
\citet{ful21} also suggest an increase in giant planet frequency beyond roughly $1 M_{\odot}$ using the 178 planets discovered by the California Legacy Survey \citep{ros21}.
Importantly, part of their planet sample overlaps with our giant planet sample in the parameter space of mass ratio and semi-major axis.
On the other hand, the planet samples in the RV studies do not include lower-mass planets beyond the snow line whereas our sample does.
Note that the RV planet samples were selected based on the planet masses while our subsamples were divided based on the mass ratios.

Simulations based on the core accretion theory also suggest that the population of massive planets increases as the host star mass grows \citep{burn21}.
In particular, at $M_{\rm host} \geq 0.5M_\odot$, giant planets are predicted to emerge and lead to the ejection of low-mass planets. 
\citet{liu19} and \citet{liu20} calculate the population of single planets around stars with masses between $0.1 \, M_\odot$ and $1 \, M_\odot$, and show that gas giant planets are more likely to exist around a massive star.
\citet{ida05} also predict that the Jupiter-mass planet frequency has peaks around G-dwarfs.
These theoretical results suggest $m>0$ for massive planets, which is consistent with our result.

On the other hand, results from the $Kepler$ telescope suggest that the frequency of sub-Neptunes at orbital periods less than 50 days is higher for M-dwarf rather than for FGK stars \citep{mul18}.
These results prefer $m < 0$ for low-mass planets in inner orbits.
This can be compared with our results for planets beyond the snowline that are $m = -0.08^{+0.95}_{-0.65}$ for the $q < 10^{-3}$ subsample and $m = 0.46^{+1.29}_{-0.98}$ for the $q < 10^{-3.5}$ subsample.
However, the uncertainties of our results in $m$ are large, and further investigation is needed.


\subsection{Prior for Planetary Event Analysis}
The results of this study are also important for the analysis of planetary events.
In microlensing event analysis, Bayesian analysis using the Galactic model as a prior has been used to obtain a posterior probability distribution of the lens mass and distance.
For planetary events, we have been making assumptions regarding the dependence of planet frequency on the host star mass and location in our Galaxy, i.e., assumptions on $m$ and $r$ in the context of this study.
A traditional assumption is $(m, r) = (0, 0)$, and it has been implicitly or explicitly assumed in many studies to date \citep[e.g.,][]{ben14, shv14, shi23}.
Some studies consider other possibilities for $m$ like $m = 1$ \citep{kos17, ish22, olm23} based on results by other techniques like RV \citep{jon10} which has a very different sensitivity region than microlensing, or based on a possible trend inferred from some high-angular resolution follow-up observation results for microlensing planets \citep{bha21}.

\citet{kos21a} imposed constraints on the $r$ value, $r = 0.2 \pm 0.4$.
Because this is consistent with $r=0$, the traditional assumption of $r=0$ was observationally justified.
On the other hand, the previous study has a large uncertainty regarding the host star mass dependence.
Our results succeeded in making more constraints, and in particular, found that $m > 0$ is preferred for microlensing planetary events with $q \simgt 10^{-3}$.
Therefore, using $m > 0$ (e.g., $m = 1$) might be a better choice than the traditional assumption of $m = 0$ for events with $q \simgt 10^{-3}$.

\subsection{Comparison with the Previous Method} \label{compare-kos21}
As we discussed in Section \ref{intro}, this study is an extension of \citet{kos21a}. As a comparison, we analyzed the same 22 planet samples used in our study by using the method by \citet{kos21a}.
Fig. \ref{comp_kos_nuno} compares the result with these two methods.
This corresponds to a comparison between the result using the one-dimensional distribution of $\mu_{\rm rel}$ for given $t_{\rm E}$ and the result using the two-dimensional distribution of $t_{\rm E}$ and $\mu_{\rm rel}$.
As expected, Fig. \ref{comp_kos_nuno} shows that using a two-dimensional distribution allows for more constraint of $m$ and $r$ values compared to using the one-dimensional distribution.

A disadvantage of the new method is that the number of samples is limited to apply proper detection efficiency as described in Section \ref{sec-pla_sample}. 
In fact, in this study, 6 planetary events from \citet{gou10} and \citet{cas12} were excluded for that reason.
On the other hand, the previous method has the advantage of easily increasing the sample size by avoiding the detection efficiency issue, and \citet{kos21a} used the 6 events mentioned above in addition to the 22 events used in this study.

Nevertheless, we were able to impose more constraint of $m = 0.50^{+0.90}_{-0.70}$ compared to $m = 0.2\pm 1.0$ by \citet{kos21a}.
This fact indicates that the two-dimensional approach is more informative than the one-dimensional approach even considering the decrease in the number of samples.
Hence, when detection efficiency is available, it is preferable to use the method described in this study as much as possible.
Note that both methods require a Galactic model, and one needs to ensure that the model is unbiased, for instance, by a sanity test as performed in Appendix {\ref{sec-model}.

\section{Conclusion}\label{sec-conclu}
We estimated the dependence of planet frequency on the host star mass and the Galactocentric distance by comparing the $(t_{\rm E}, \mu_{\rm rel})$ distribution of the 22 microlensing planetary events from \citetalias{suz16} with the one expected from the Galactic model.
By assuming the power law $P_{\rm host} \propto M_{\rm L}^m \times R_{\rm L}^r$ as the planet-hosting probability, we estimated $r = 0.10^{+0.51}_{-0.37}$ and $m = 0.50^{+0.90}_{-0.70}$.
We also divided our sample into subsamples by the mass ratios and found that the giant planet sample with $q \simgt 10^{-3}$ prefers $m>0$ whereas $m$ is consistent with 0 for the lower-mass ratio samples.
It suggests that massive planets are more likely to exist around more massive stars. 
On the other hand, there is no significant preference in either positive or negative $r$, i.e., no large dependence of planet frequency on the Galactocentric distance, which is consistent with the result of \citet{kos21a}.

The analysis method of this study and \citet{kos21a} can be used for planet samples from other microlensing survey projects.
The Korea Microlensing Telescope Network \citep[KMTNet;][]{kim16} has operated their microlensing survey since 2016 and more than 200 planets have already been detected.
The PRime-focus Infrared Microlensing Experiment (PRIME) began their survey toward the Galactic bulge and center in 2023 \citep{kon23,yam23}.
PRIME is expected to discover $42-50$ planets per year \citep{kon23}.
The Nancy Grace Roman Space Telescope is planned to launch in late 2026 \citep{spe15} and a total of $\sim 1400$ planets is expected to be discovered \citep{pen19}.
A similar analysis with the planet sample by these surveys can further constrain the dependence of planet frequency on the host star mass and the location in our Galaxy.

\begin{acknowledgments}
KN was supported by the Scholarship of the Graduate School of Science of ``Osaka University Foundation for the Future" for overseas research activities in 2023, JSPS Core-to-Core Program (grant number: JPJSCCA20210003), and, Ono Scholarship Foundation for a public interest incorporated foundation.
DS was supported by JSPS KAKENHI grant No. JP19KK082.
We are grateful to Kento Masuda and Ryusei Hamada for their insightful discussions.

\end{acknowledgments}

\appendix
\section{Validation of Method by Mock Data Analysis}\label{sec-mock-test}
As described in Section \ref{sec-method}, we conduct mock data simulations to validate our method.
We adopt a certain ($m$,$r$) value and generate 50 mock planetary event samples with weights of $\Gamma_{\rm host}(t_{\rm E}, \mu_{\rm rel}|m,r)\times \epsilon$. This sample can be regarded as a sample of actually observed planetary events in the virtual galaxy which has a specific value of $(m,r)$ and we know this specific value.
Therefore, if the analysis method is correct, it is expected that we can reproduce the actual values of $(m,r)$ by analyzing these mock planetary events.
We produced mock data with nine combinations of $m=-1,0,1$, and $r=-1,0,1$, and analyzed these artificial planetary events.

Fig. \ref{mock_tE_murel} (a) shows the distribution of $t_{\rm E}$ and $\mu_{\rm rel}$ under each $(m, r)$ value, and Fig. \ref{mock_tE_murel} (b) shows the result of the likelihood analysis for the mock data indicated by dots in Fig. \ref{mock_tE_murel} (a).
It can be seen that the correct $m$ and $r$ values are well reproduced in each analysis regardless of the true $(m, r)$ values although the strength of the $(m,r)$ correlation differs depending on the true $(m,r)$ values.\\

\begin{figure*}
\gridline{\fig{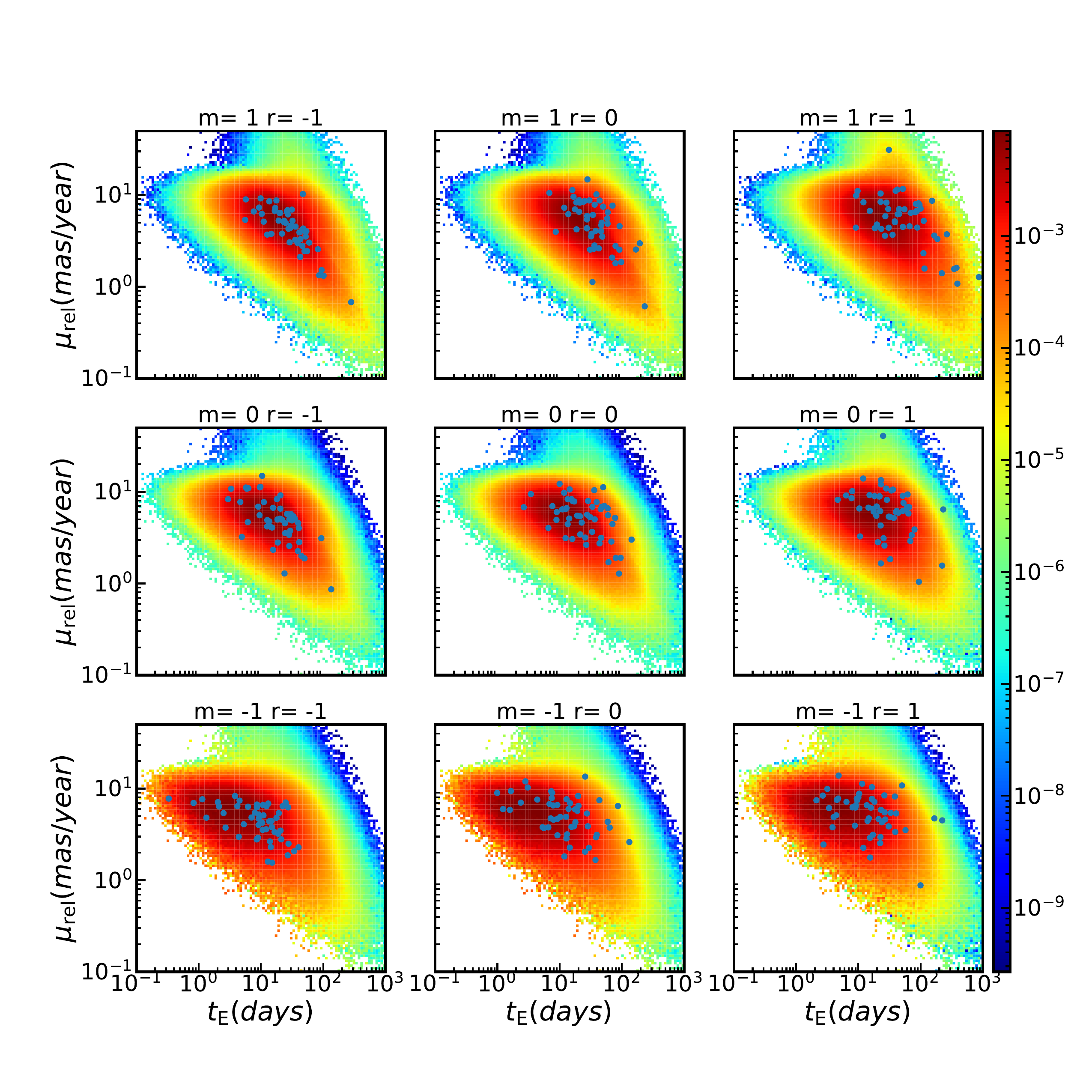}{0.5\textwidth}{(a)}
          \fig{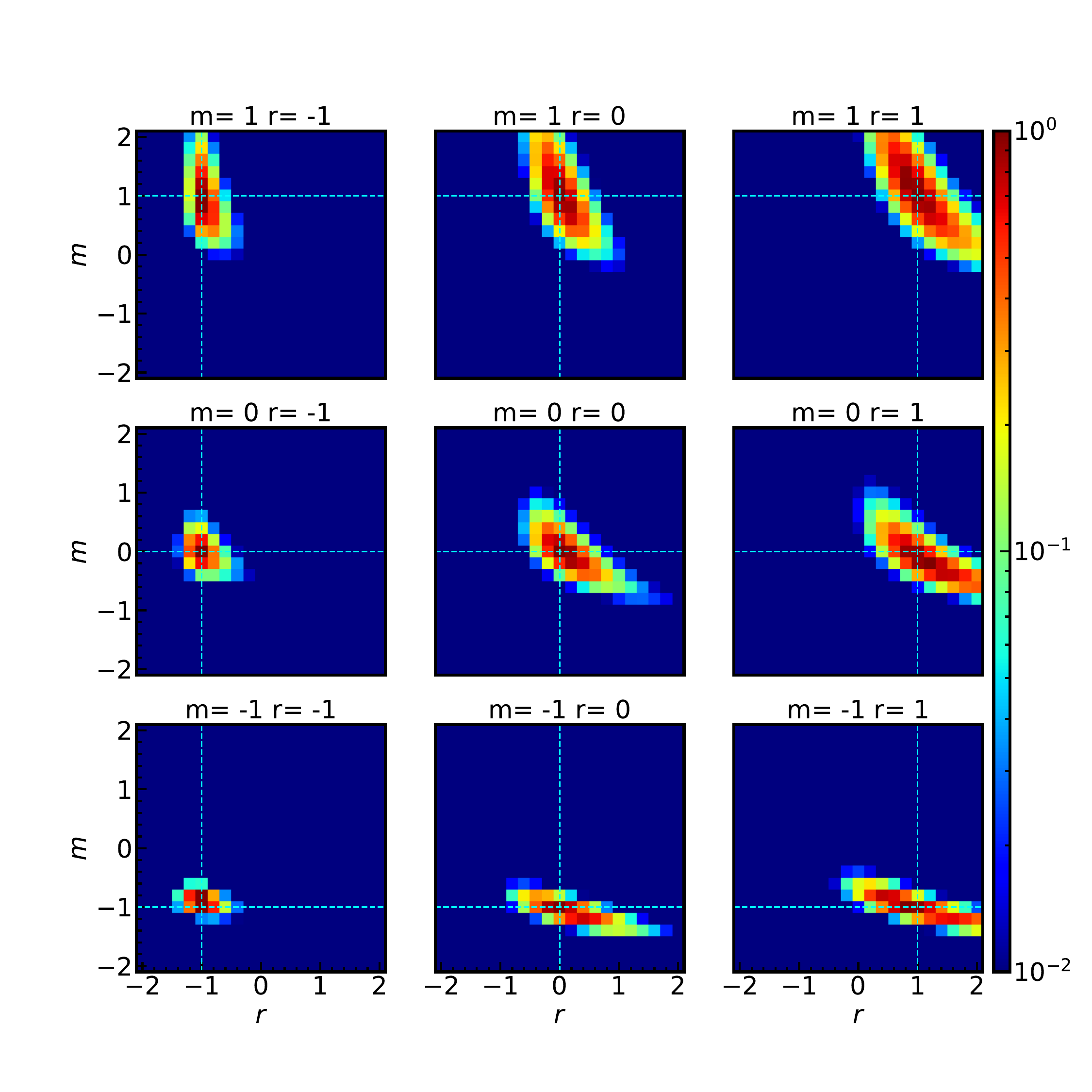}{0.5\textwidth}{(b)}}
\caption{
(a) Two-dimensional distribution of $t_{\rm E}$ and $\mu_{\rm rel}$. The blue dots show the 50 samples used in this mock data analysis. (b) Result of the mock data analysis. Each panel has a different correct $(m,r)$ value and the intersection of the dotted lines shows its correct value. 
Each analysis is based on 50 artificially generated samples, weighted according to their respective $(m,r)$ value.\label{mock_tE_murel}}
\end{figure*}


\section{Verification of Galactic model by MOA-II FSPL sample} \label{sec-model}
Although our method was validated by the mock data analysis in Appendix \ref{sec-mock-test}, the simulations using mock data are not sufficient to truly justify the results from this study, because the same Galactic model was used both to generate the mock data and to calculate the likelihood.
Since the real data are generated following the real distribution of our Galaxy, the validity of the Galactic model needs to be verified.

In this section, we compare the $(t_{\rm E}, \mu_{\rm rel})$ distribution predicted by the Galactic model with the distribution of the finite-source point-lens (FSPL) event sample from the MOA-II 9-yr Galactic bulge survey \citepalias{kos23} to evaluate the amount of bias in the Galactic model that would affect our measurement of $(m, r)$.
Because the FSPL sample should reflect the distribution of random stars in our Galaxy, their $(t_{\rm E}, \mu_{\rm rel})$ distribution can be fairly compared with the predicted distribution by the Galactic model if the detection efficiency is properly taken into account.

For the comparison in this section, we use a slightly modified version of Eq. (\ref{eq-fobs}) for the model predicted distribution, i.e.,
\begin{equation}
f_{\rm FSPL} (t_{\rm E}^{\rm (obs)}, \mu_{\rm rel}^{\rm (obs)}| m_{\rm bias},r_{\rm bias}) = \notag \int dt_{\rm E}^{,} d\mu_{\rm rel}^{,} \Bigl[k\left({t_{\rm E}^{\rm (obs)}, \mu_{\rm rel}^{\rm (obs)}};t_{\rm E}^{,},\mu_{\rm rel}^{,}\right)\notag \Gamma_{\rm bias}\left(t_{\rm E}^{,}, \mu_{\rm rel}^{,}|m_{\rm bias},r_{\rm bias}\right) \epsilon_{\rm FSPL} \left(t_{\rm E}^{,}, \mu_{\rm rel}^{,}\right)\Bigr],
\label{eq-fobsFS} 
\end{equation}
where $(t_{\rm E}^{\rm (obs)}, \mu_{\rm rel}^{\rm (obs)})$ are from the MOA-II 9-yr FSPL sample described in Section \ref{sec-FSPL}, $\epsilon_{\rm FSPL} (t_{\rm E}, \mu_{\rm rel})$ is the detection efficiency corresponding to the sample described in Section \ref{FSPL_det_eff}.
$\Gamma_{\rm bias}\left(t_{\rm E}, \mu_{\rm rel}|m_{\rm bias},r_{\rm bias}\right)$ is defined as 
\begin{equation}
\Gamma_{\rm bias}\left(t_{\rm E}, \mu_{\rm rel}|m_{\rm bias},r_{\rm bias}\right)\propto \notag \int dM_{\rm L}dR_{\rm L} \, \Gamma_{\rm FSPL}( t_{\rm E},\mu_{\rm rel},  M_{\rm L}, R_{\rm L}) M_{\rm L}^{m_{\rm bias}} R_{\rm L}^{r_{\rm bias}}, 
\label{eq-Gbias} 
\end{equation}
where $\Gamma_{\rm FSPL}$ is the event rate for FSPL events calculated by the Galactic model and $\Gamma_{\rm FSPL} \propto \Gamma_{\rm all}\, \theta_{\rm E}^{-1}$. $m_{\rm bias}$ and $r_{\rm bias}$ are the parameters to quantify the bias level in the Galactic model and $(m_{\rm bias}, r_{\rm bias}) = (0, 0)$ corresponds to no bias. We evaluate $(m_{\rm bias}, r_{\rm bias})$ for the \citet{kos21b} Galactic model in Section \ref{sec-result-FSPL}.

\subsection{MOA-II 9-yr FSPL sample} \label{sec-FSPL}
\citetalias{kos23} systematically analyzed the MOA-II Galactic bulge survey data during the 9 years from 2006 to 2014 and selected $\sim 3500$ single-lens events.
There are 13 FSPL events in the MOA-II 9-yr sample where the finite source effect \citep{alc97, yoo04} was detected, and both $t_{\rm E}$ and $\mu_{\rm rel}$ were measured thanks to the effect.
Two of the FSPL events are free-floating planet candidates with $t_{\rm E} < 0.5 ~{\rm days}$, and modeling their distribution requires an additional part of the mass function for a planetary mass range that is irrelevant to our sample of the \citetalias{suz16}'s events with $t_{\rm E} > 2~{\rm days}$.
Therefore, we do not use the two events and consider the remaining 11 FSPL events. 
This corresponds to applying an additional selection criterion of $\theta_{\rm E} > 0.03~{\rm mas}$ to the 9-yr sample in addition to the original selection criteria applied by \citetalias{kos23}.
This additional selection criterion allows us to avoid considering events with extremely small $\theta_{\rm E}$ and to erase the $\mu_{\rm rel}$ dependency from the detection efficiency for single-lens events, $\epsilon_{\rm SL} (t_{\rm E})$, which is defined below in Section \ref{FSPL_det_eff}.

The black dots in the left panel of Fig.\ref{FSPL_tE_murel} show the $(t_{\rm E}, \mu_{\rm rel})$ distribution of the selected 11 FSPL events.

\begin{figure*}
\begin{center}
\includegraphics[width=8cm]{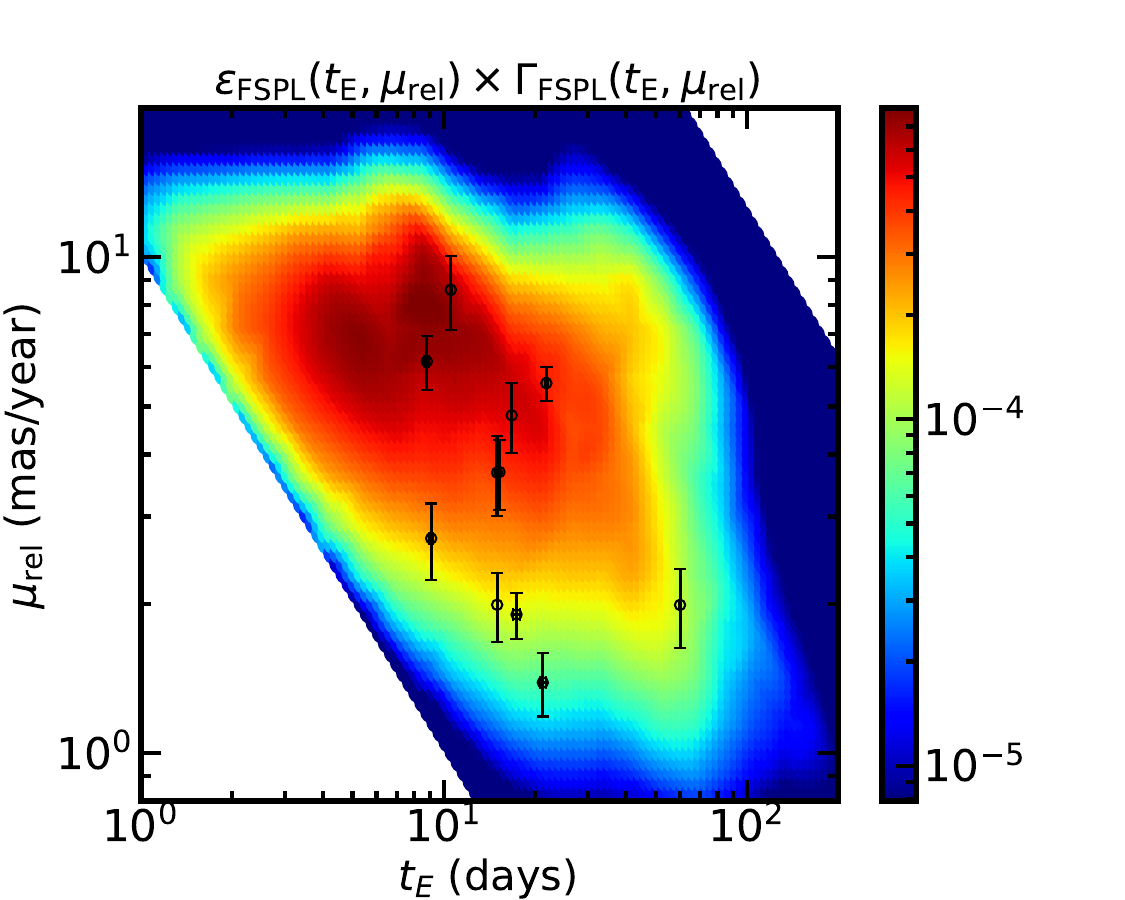}
\includegraphics[width=8cm]{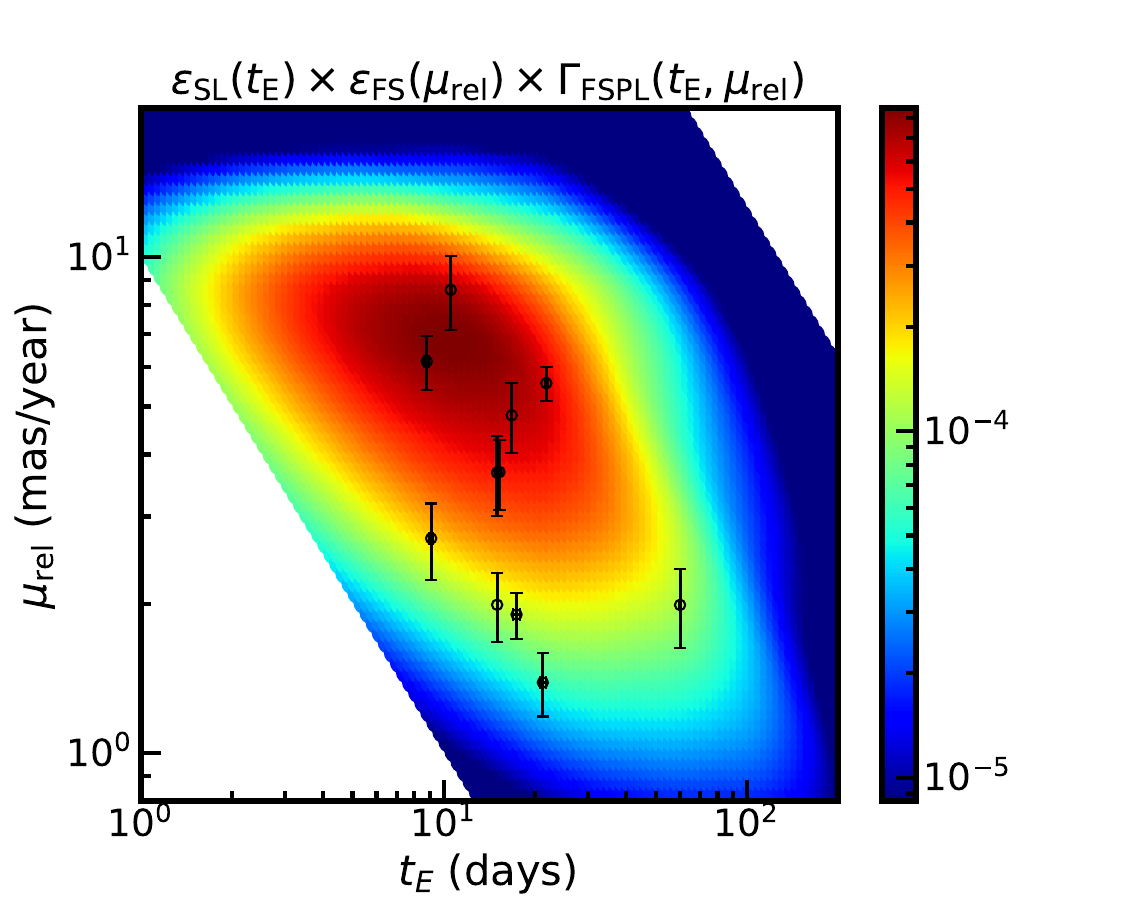}
\caption{
Comparison of the $(t_{\rm E}, \mu_{\rm rel})$ distributions for the MOA-II 9-yr FSPL sample between the observations (black points) and the prediction by the Galactic model combined with the detection efficiency (color maps) when $(m_{\rm bias}, r_{\rm bias}) = (0, 0)$.
The left panel calculates the detection efficiency, $\epsilon_{\rm FSPL} (t_{\rm E}, \mu_{\rm rel})$, without the separable assumption (Eq. \ref{eq:ep_FS}), whereas the right panel calculates the one with the separable assumption (Eq. \ref{eq:ep_FS_sep}).
\label{FSPL_tE_murel}}
\end{center}
\end{figure*}

\subsection{Detection Efficiency}\label{FSPL_det_eff}
\citetalias{kos23} performed image-level simulations of $6.4 \times 10^7$ artificial events to calculate the detection efficiency as a function of $(t_{\rm E}, \theta_{\rm E})$ which is equivalent to the detection efficiency as a function of $(t_{\rm E}, \mu_{\rm rel})$ because $\theta_{\rm E} = \mu_{\rm rel} \, t_{\rm E}$.
However, their original detection efficiency is for their sample of $\sim 3500$ single-lens events including both PSPL and FSPL events, and not suitable for our sample of the 11 FSPL events.
Therefore, we recalculate the detection efficiency for our sample, $\epsilon_{\rm FSPL} (t_{\rm E}, \mu_{\rm rel})$, using their simulation results by the following procedure.

Their simulated artificial events were distributed uniformly in $0 < u_0 < 1.5$, where $u_0$ is the impact parameter in units of $\theta_{\rm E}$.
To define the detection efficiency against FSPL events that follow $\Gamma_{\rm FSPL} \propto \Gamma_{\rm all} \, \theta_{\rm E}^{-1}$, we first limit to their artificial events with $0 < z_0 < 5$, where $z_0 \equiv u_0/\rho$ and $\rho = \theta_*/\theta_{\rm E}$ is the size of the angular source radius $\theta_*$ in units of $\theta_{\rm E}$.
Then, we further limit the artificial events to those with $\theta_{\rm E} > 0.03~{\rm mas}$ as described in Section \ref{sec-FSPL}.
The remaining artificial events are used as the total number of valid simulated events, $N_{{\rm sim}, z_0}$. Note that event counts here are done after considering the weight for each event based on its event rate given by Eq. (13) of \citetalias{kos23}.

The next step is to count the number of events that pass the selection criteria for the FSPL events.
There are two requirements to be selected as an FSPL event in \citetalias{kos23}: the first is to pass the selection criteria listed in Table 2 of \citetalias{kos23} and be selected as a single lens event, and the second one is to have a significant $\Delta \chi^2$ value between the best-fit PSPL and FSPL models (see Section 8 of \citetalias{kos23} for the detail).
We apply the same two-step criteria and count the number of events that pass the first step as $N_{\rm det, SL}$ and the ones that also pass the second step as $N_{\rm det, FSPL}$.

Ideally, the desired detection efficiency $\epsilon_{\rm FSPL} (t_{\rm E}, \mu_{\rm rel})$ can be simply calculated by 
\begin{align}
\epsilon_{\rm FSPL} (t_{\rm E}, \mu_{\rm rel}) = \frac{N_{\rm det, FSPL} (t_{\rm E}, \mu_{\rm rel})}{N_{{\rm sim}, z_0} (t_{\rm E}, \mu_{\rm rel})}, \label{eq:ep_FS}
\end{align}
where $N_{\rm det, FSPL} (t_{\rm E}, \mu_{\rm rel})$ and $N_{{\rm sim}, z_0} (t_{\rm E}, \mu_{\rm rel})$ are subsamples of $N_{\rm det, FSPL}$ and $N_{{\rm sim}, z_0}$ in a grid of $(t_{\rm E}, \mu_{\rm rel})$, respectively.
However, the number of $N_{\rm det, FSPL}$ in each grid of $(t_{\rm E}, \mu_{\rm rel})$ is too small to have a precise value of 
$\epsilon_{\rm FSPL} (t_{\rm E}, \mu_{\rm rel})$ 
because the \citetalias{kos23}'s simulation was not optimized for FSPL events.
Therefore, we assume that $\epsilon_{\rm FSPL} (t_{\rm E}, \mu_{\rm rel})$ is separable as a product of two single variable functions, i.e., 
\begin{align}
\epsilon_{\rm FSPL} (t_{\rm E}, \mu_{\rm rel}) \simeq \epsilon_{\rm SL} (t_{\rm E}) \, \epsilon_{\rm FS} (\mu_{\rm rel}), \label{eq:ep_FS_sep}
\end{align}
where 
\begin{align}
\epsilon_{\rm SL} (t_{\rm E}) &= \frac{N_{\rm det, SL} (t_{\rm E})}{N_{{\rm sim}, z_0} (t_{\rm E})},
\end{align}
and
\begin{align}
\epsilon_{\rm FS} (\mu_{\rm rel}) &= \frac{N_{\rm det, FSPL} (\mu_{\rm rel})}{N_{\rm det, SL} (\mu_{\rm rel})}.
\end{align}
Eq. (\ref{eq:ep_FS_sep}) gives us a more precise $\epsilon_{\rm FSPL} (t_{\rm E}, \mu_{\rm rel})$ distribution than Eq. (\ref{eq:ep_FS}) because it enables us to use the numbers of $N_{\rm det, FSPL}$ distributed in one-dimensional bins of $\mu_{\rm rel}$ instead of the numbers distributed in two-dimensional grids of $(t_{\rm E}, \mu_{\rm rel})$.

The separable assumption of $\epsilon_{\rm FSPL} (t_{\rm E}, \mu_{\rm rel}) = \epsilon_{\rm SL} (t_{\rm E}) \, \epsilon_{\rm FS} (\mu_{\rm rel})$ is reasonable because the detection efficiency for single lens events, $\epsilon_{\rm SL}$, only depends on $t_{\rm E}$ for events with $\theta_{\rm E} > 0.03~{\rm mas}$ as shown in Figure 7 of \citetalias{kos23}.
Also, the detection efficiency for the finite source effect depends on the number of data points taken during the source radius crossing time, $t_* = \theta_*/\mu_{\rm rel}$.
Because the angular source radius $\theta_*$ is independent of $t_{\rm E}$, the detection efficiency for the finite source effect only depends on $\mu_{\rm rel}$.

The color maps in Fig. \ref{FSPL_tE_murel} show the model-predicted $(t_{\rm E}, \mu_{\rm rel})$ distributions, i.e., $\Gamma_{\rm FSPL} (t_{\rm E}, \mu_{\rm rel}) \times \epsilon_{\rm FSPL} (t_{\rm E}, \mu_{\rm rel})$, where the left panel shows the one with $\epsilon_{\rm FSPL} (t_{\rm E}, \mu_{\rm rel})$ calculated using Eq. (\ref{eq:ep_FS}) and the right panel shows the one with $\epsilon_{\rm FSPL} (t_{\rm E}, \mu_{\rm rel})$ calculated using Eq. (\ref{eq:ep_FS_sep}).
As expected, the right panel shows a much smoother distribution than the left panel.
At the same time, the two distributions look like they represent a similar distribution, indicating that the separable assumption is valid, at least to a good approximation.\\

\begin{figure}
\centering
\includegraphics[width=9cm]{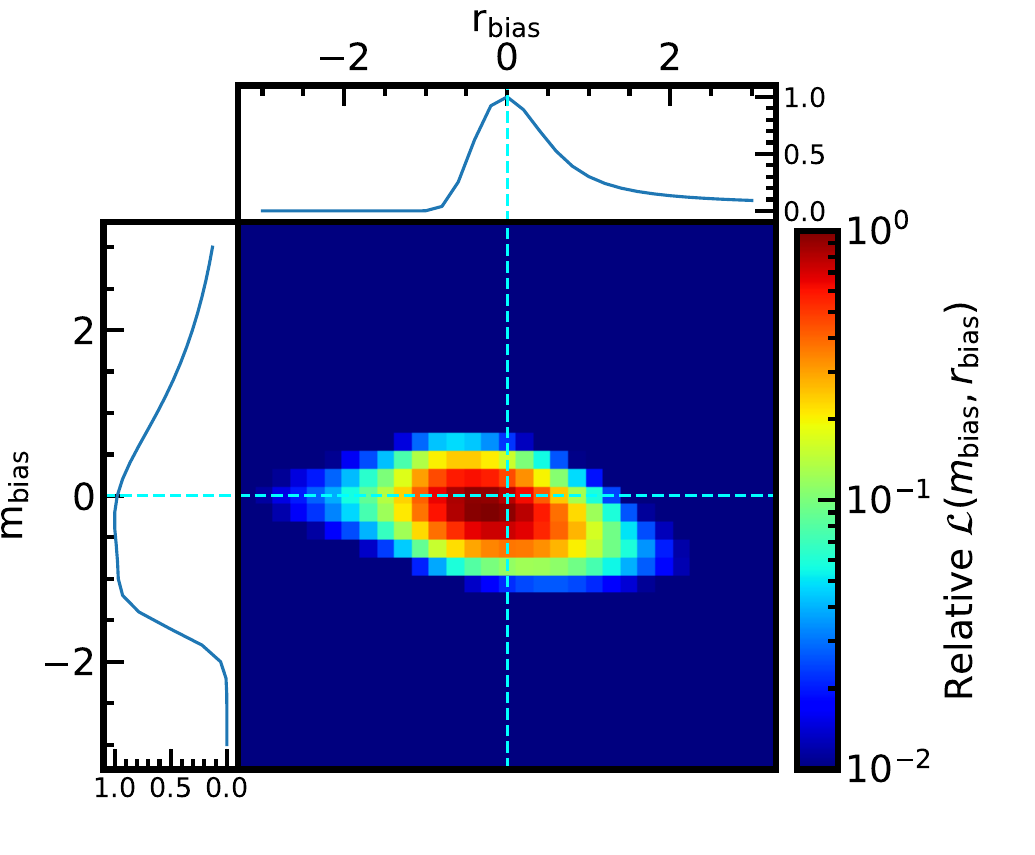}
\caption{
Relative likelihood distribution of $(m_{\rm bias}, r_{\rm bias})$ calculated by Eq. (\ref{eq:L_bias}) for the Galactic model by \citet{kos21b} when compared with the MOA-II 9-yr FSPL sample.
The top panel shows a relative probability distribution of $r_{\rm bias}$ integrated over $-3< m_{\rm bias}<3$ uniformly, and the side panel shows a relative probability distribution of $m_{\rm bias}$ integrated over $-3<r_{\rm bias}<3$ uniformly.
}
\label{FSPL_result}
\end{figure}

\subsection{Sanity Test for the Galactic Model} \label{sec-result-FSPL}
Fig. \ref{FSPL_tE_murel} shows the comparison of the observational $(t_{\rm E}, \mu_{\rm rel})$ distribution from the MOA-II 9-yr FSPL sample (black dots) with the one from the Galactic model with $(m_{\rm bias}, r_{\rm bias}) = (0, 0)$ (color map).
It shows a good agreement between the observations and the model, which implies that the Galactic model by \citet{kos21b} is not significantly biased.
To quantify this, we calculate the likelihood distribution of $(m_{\rm bias}, r_{\rm bias})$ given by 
\begin{align}
{\cal L} (m_{\rm bias}, r_{\rm bias}) = \prod_{i} f_{\rm FSPL} (t_{{\rm E},i}^{\rm (obs)}, \mu_{{\rm rel},i}^{\rm (obs)}| m_{\rm bias}, r_{\rm bias}), \label{eq:L_bias}
\end{align}
and the result is shown in Fig. \ref{FSPL_result}.
Fig. \ref{FSPL_result} shows that the likelihood is distributed around $(m_{\rm bias}, r_{\rm bias}) = (0,0)$.
The best grid is at $(m_{\rm bias}, r_{\rm bias}) = (-0.2,-0.2)$ and the likelihood at $(m_{\rm bias}, r_{\rm bias}) = (0,0)$ is $0.80$ relative to the best grid value.
The median and $1\sigma$ uncertainty values are $m_{\rm bias}= -0.27^{+0.33}_{-0.32}$ and $r_{\rm bias}= -0.32^{+0.65}_{-0.61}$.

The fact that the likelihood distribution is consistent with $(m_{\rm bias}, r_{\rm bias}) = (0,0)$ means that the Galactic model by \citet{kos21b} would not cause a significant bias in our estimation on $(m, r)$, and we can securely use the model in our analysis in Section \ref{sec-planet}.
The same test can be used for any other Galactic models to test their validity. 
\\

\end{document}